\documentclass[12pt,preprint]{aastex}

\shortauthors{Zheng \& Wang}
\begin{document}

\title{Relativistic Outflows in two quasars in the Chandra Deep Field South
}
\author{Z. Y. Zheng\altaffilmark{1},
and
J. X. Wang\altaffilmark{1}
}
\altaffiltext{1}{Center for Astrophysics, University of Science and Technology of China, Hefei, Anhui 230026, P. R. China; zhengzy@mail.ustc.edu.cn and jxw@ustc.edu.cn.}
\vspace{0.1cm}

\begin{abstract}
In this paper, we provide new 1 Ms $Chandra$ ACIS spectra of two quasars
in the Chandra Deep Field South (CDF-S), which were previously reported to
show strong and extremely blueshifted X-ray emission/absorption line features
in previous 1 Ms spectra, with outflowing bulk velocity 
$v\sim$0.65-0.84c. In the new 1 Ms spectra, the relativistic blueshifted line feature is solidly confirmed in CXO CDFS 
J033225.3-274219 (CDFS 46, $z$ = 1.617), and marginally visible in 
CXO CDFS J033260.0-274748 (CDFS 11, $z$ = 2.579), probably due to the increased
Chandra ACIS background in the new 1 Ms exposure.
The new data rule out the possibility (though very tiny already based on the
old 1Ms data) that the two sources were selected to be unusual due to noise
spikes in the spectra.
The only likely interpretation
is extremely blueshifted iron absorption/emission line or absorption edge due to relativistic outflow.
We find that the rest frame emission line center in CDFS 46 marginally decreased 
from 16.2 keV to 15.2 keV after 7 years.
The line shift can be due to either decreasing outflowing velocity or 
lower ionization level. Including the two quasars reported in this paper, we collect from 
literature a total of 7 quasars showing blueshifted emission or 
absorption line feature with $v\geq0.4c$ in X-ray spectra, and discuss 
its connection to jet and/or BAL (broad absorption line) outflow.

\end{abstract}

\section{Introduction}

It's well known that quasars interact with their
environments through collimated jets and outflowing winds, which is 
required as the feedback (from the central supermassive black hole to
the host galaxy) for the the M-$\sigma_*$ relation (Gebhardt et al. 2000).
The collimated jets and outflowing winds are also natural diagnostics to
study the gas flow patterns in the innermost regions of black holes,
the source geometry and energy generation mechanism (e.g., see Cappi 2006).

Intrinsic outflowing systems may be categorized by their outflowing velocities.
The jets are often seen in radio-loud quasars with velocities very close
to the speed of light
(except for shocks and lobe-dominated sources, which are interacted with
the surrounding medium and appear to be sub-relativistic, e.g., Kellermann
et al. 2004). The broad absorption lines (BALs) in
UV/optical band are blueshifted with velocities
up to 0.2c, presenting in the spectra of 
10$\sim$20\% quasars. 
The BAL outflow can be seen in the X-ray spectra at even higher velocity.
Chartas et al. (2002, 2003, 2007a, 2007b) discovered very broad absorption lines
with outflow velocity of 0.4-0.67c in three gravitational lensed radio quiet BAL quasars (PG 1115$+$080, H 1413$+$117, and APM 08279$+$5255; see table 1 for details), which are the only BALs reported with relativistic outflows ($v\geq0.4c$) in their X-ray spectrum. 

Meanwhile, relativistic outflows with $v\sim0.6-0.8c$ are also seen in the
X-ray spectra of four non-BAL quasars. Yaqoob et al. (1998) reported
blueshifted O$_{VII}$ emission line in the ASCA spectrum of PKS 0637-752 
with outflow velocity of 0.77c. The line is however invisible in the
Chandra exposure obtained $\sim$ 2 years later, which instead detected
an extended X-ray jet in this core-dominated radio-loud quasar.
Yaqoob et al. (1999) reported the detection of blueshifted iron K emission line
in radio-loud quasar PKS 2149-306 with velocity of 0.71c (see Table 1) in 
the ASCA spectrum. Such feature is also visible in following up XMM
spectrum (see Fig. 2 of Page et al. 2005), and our independent analysis gives a confidence
level of 99\% in XMM PN spectrum.
With the 1 Ms Chandra exposure on Chandra Deep Field South (CDF-S), Wang et al.
(2003; 2005) reported the detection of strong blueshifted iron K line features
with $v\sim$0.65-0.84c, in CXO CDFS J033225.3-274219 (CDFS 46, $z$ = 1.617, 
emission line, radio-loud) and CXO CDFS J033260.0-274748 (CDFS 11, 
$z$ = 2.579, absorption line, radio-quiet) respectively.
The blueshift velocities of the three emission lines reported are far larger than
that could be produced by the rotational velocities of the accretion disk, and
can only be due to outflow (see Wang et al. 2003).

In this paper, we report new 1 Chandra Ms observation\footnote{The $Chandra$ Deep Field South 
(CDF-S, e.g., Giacconi et al. 2002; Rosati et al. 2002) now has a newly 1 Ms exposure
(from 2007 September to 2007 November), and the data has been released. See 
http://cxc.harvard.edu/cda/whatsnew.html\#CDFS2000-2007}  of the two
 CDF-S sources with relativistic outflows. 
Their multi-band properties are described in details in 
Wang et al. (2003; 2005). The X-ray data and spectral fitting are reported in \S2
and the nature of the blueshifted features are discussed in \S3.
Throughout the paper we assume a cosmological model with $H_0 = 70$ km s$^{-1}$
 Mpc$^{-1}$,  $\Omega_m$ = 0.3, and $\Omega_{\Lambda}$ = 0.7.

\section{The Data and X--ray spectral fitting}

The newly released 1~Ms $Chandra$ exposure on CDF-S was composed of 12 individual ACIS observations obtained from September to November, 2007. Each observation 
was filtered to include only standard \emph{ASCA} event grades. Cosmic ray 
afterglows, ACIS hot pixels and bad pixels were removed, and high background 
intervals were subtracted. All exposures were then added to produce a combined
 event file with net exposure of 954 ks. The two sources, CDFS 46 and CDFS 11,  
 $\sim$ 6$\arcmin$  and $\sim$ 7$\arcmin$  from the center of the field, respectively, 
 were covered by all 12 exposures.  
 The X-ray count rates of CDFS 46 and CDFS 11 extracted 
 from 12 exposures are examined and no significant fluctuations ($>$ 3$\sigma$) 
 are seen. 

We extract the \emph{Chandra} ACIS-I X-ray spectra of CDFS 46 and 
CDFS 11 from the 95\% encircled-energy radius $r_S$ of the ACIS point-spread 
function at each source's position, and the background spectra are extracted
from the annulus at the same position with inner and outer radius of $1.2r_S$
and $2.4r_S$.
The X-ray telescope response and CCD  ACIS-I instrument
 response were generated for each single \emph{Chandra} observation, 
 and then summed together with corresponding exposure times as 
 weighting factor.  The final time-weighted response files were used 
 for spectral analysis. The net counts
of CDFS 46 and CDFS 11 are 375 and 963 in the 0.5--2.0 keV band, 170 and 451 in
the 2.0--9.0 keV band. During the fit, we use the \emph{C}-statistics 
 (Cash 1979; Nousek \& Shue 1989), which perform better than the
 $\chi^2$ analysis for spectra with low detected counts. 
 We use XSPEC version 12.0 to perform the spectral fitting. All the spectral
 fitting was done in the energy band 0.5--9.0 keV, and all the statistical 
 errors quoted in this paper are at the 90\% confidence level for one 
 interesting parameter. 
 
The spectra of both CDFS 46 and CDFS 11 were binned to have at least one count per bin, and were first fitted with a simple power
 law plus a neutral absorber in the quasar frame.  A galactic neutral hydrogen 
 absorption column of 8 $\times$ 10$^{19}$ cm$^{-2}$ (Dickey \& Lockman
 1990) was also included. The fitting results of CDFS 46 are presented in Table 2.

\begin{figure}
\plotone{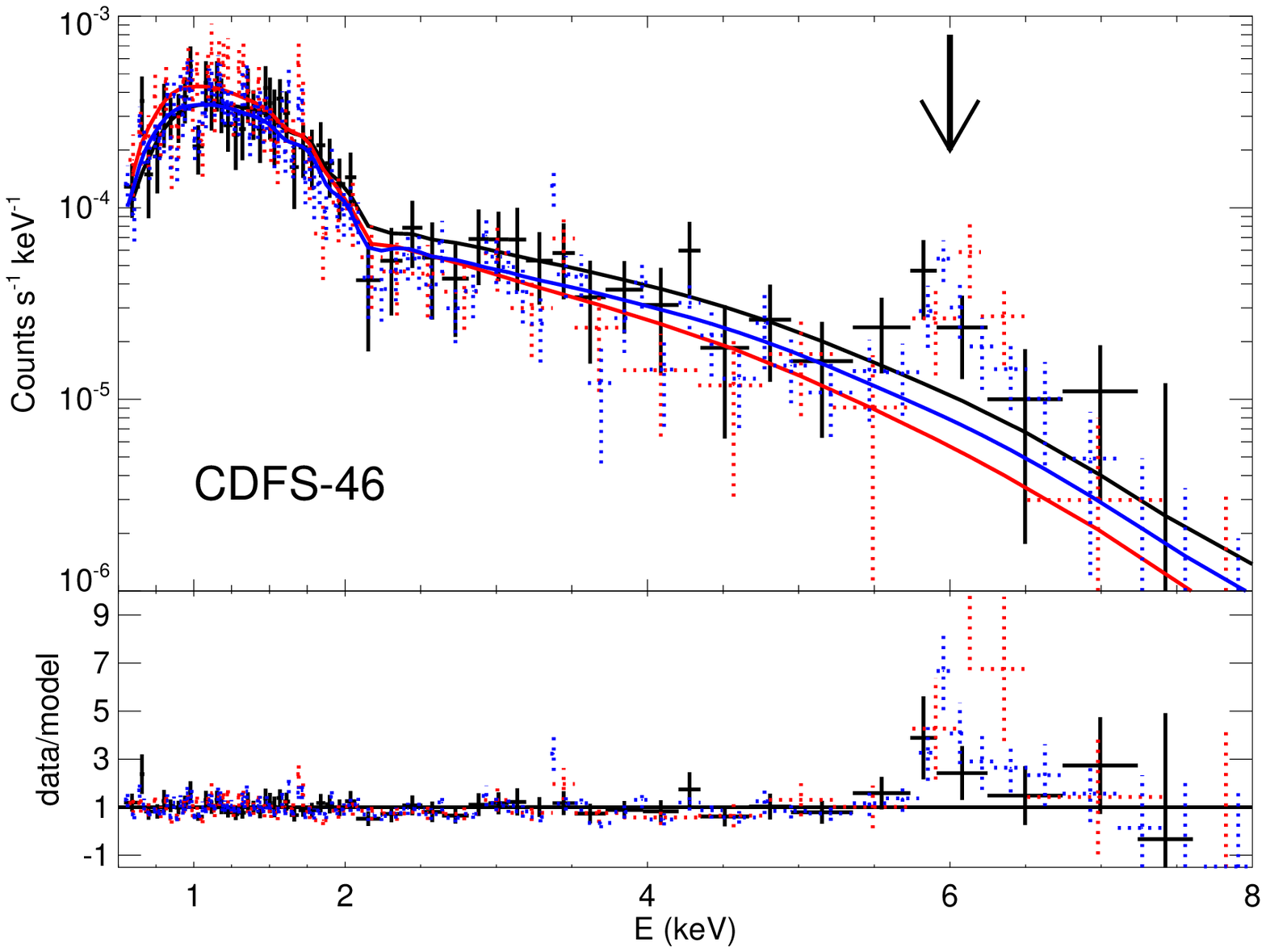}
\caption{The new 1 Ms (black), old 1 Ms (red) and combined 2 Ms (blue) spectra of CDFS 46. The
unusual emission line features at $\sim6.0$ keV (observed frame) are visible in all three spectra. The spectra were rebinned at least 10 counts per bin for display purpose only.
}
\label{spec46}
\end{figure}

\begin{figure}
\plotone{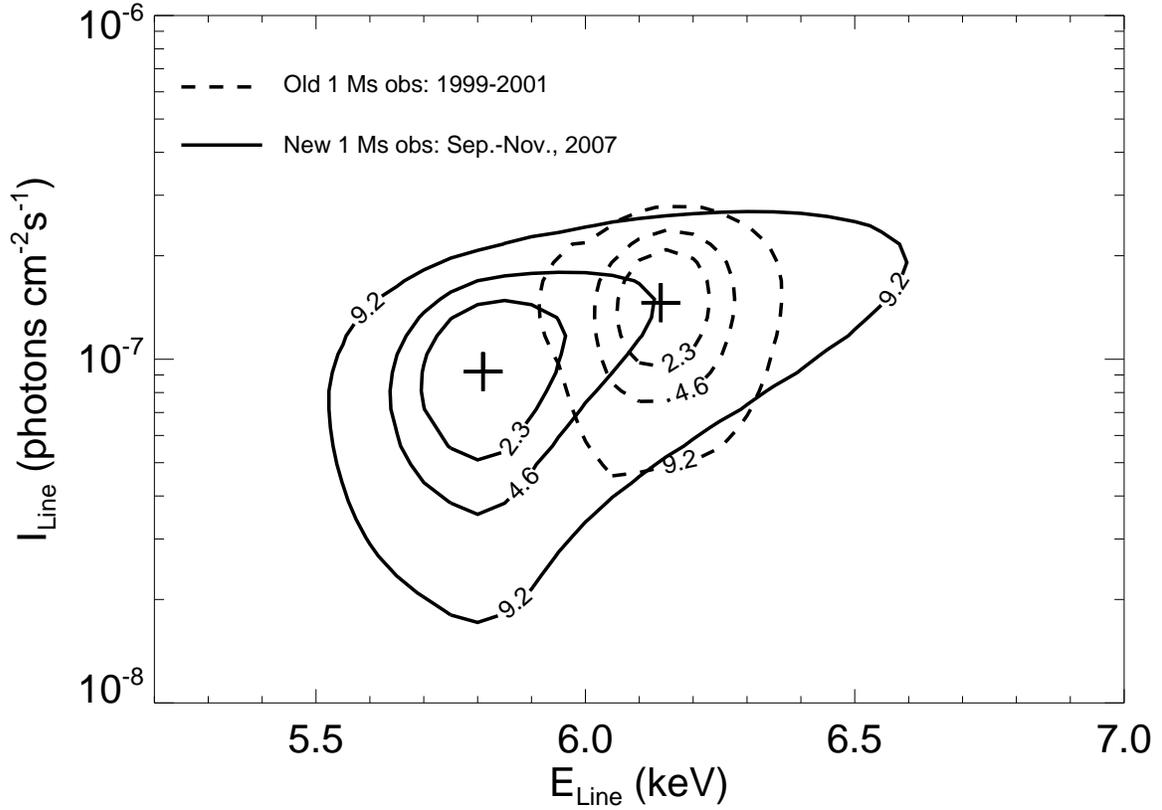}
\caption{
Contours (68\%, 90\% and 99\%) for the blueshifted line center  E$_c$ vs. 
intensity I$_{Line}$ in CDFS 46 of the old 1 Ms (dashed) exposure and the new 1 Ms (solid) exposure, both in the observed frame. The numbers on the contours are the two-dimensional delta fit C-statistic, $\Delta C$=-2.3, -4.6, and -9.2, equal to confidence levels of 68\%, 90\%, and 99\%.
}
\label{emission}
\end{figure}

In Fig. \ref{spec46} we plotted Chandra spectra of CDFS
46, for old, new 1 Ms exposure, and a combined 2 Ms exposure respectively.
In the new 1 Ms exposure, the emission line feature at around 6 keV (observed
frame) is obvious.
A single Gaussian was added to our continuum model.
With three more free parameters ($E_c$, $\sigma$, and $I$), the fit was 
significantly improved ($\Delta C$ =  -18.2) , indicating a confidence level
of 99.99\% based on F-test. 
However, comparing with the old 1 Ms exposure, we find marginal shift of
the rest frame line center 
from 16.2$^{+0.4}_{-0.3}$ to $15.2^{+0.3}_{-0.4}$ keV (see Fig. \ref{emission}).

\begin{figure}
\plotone{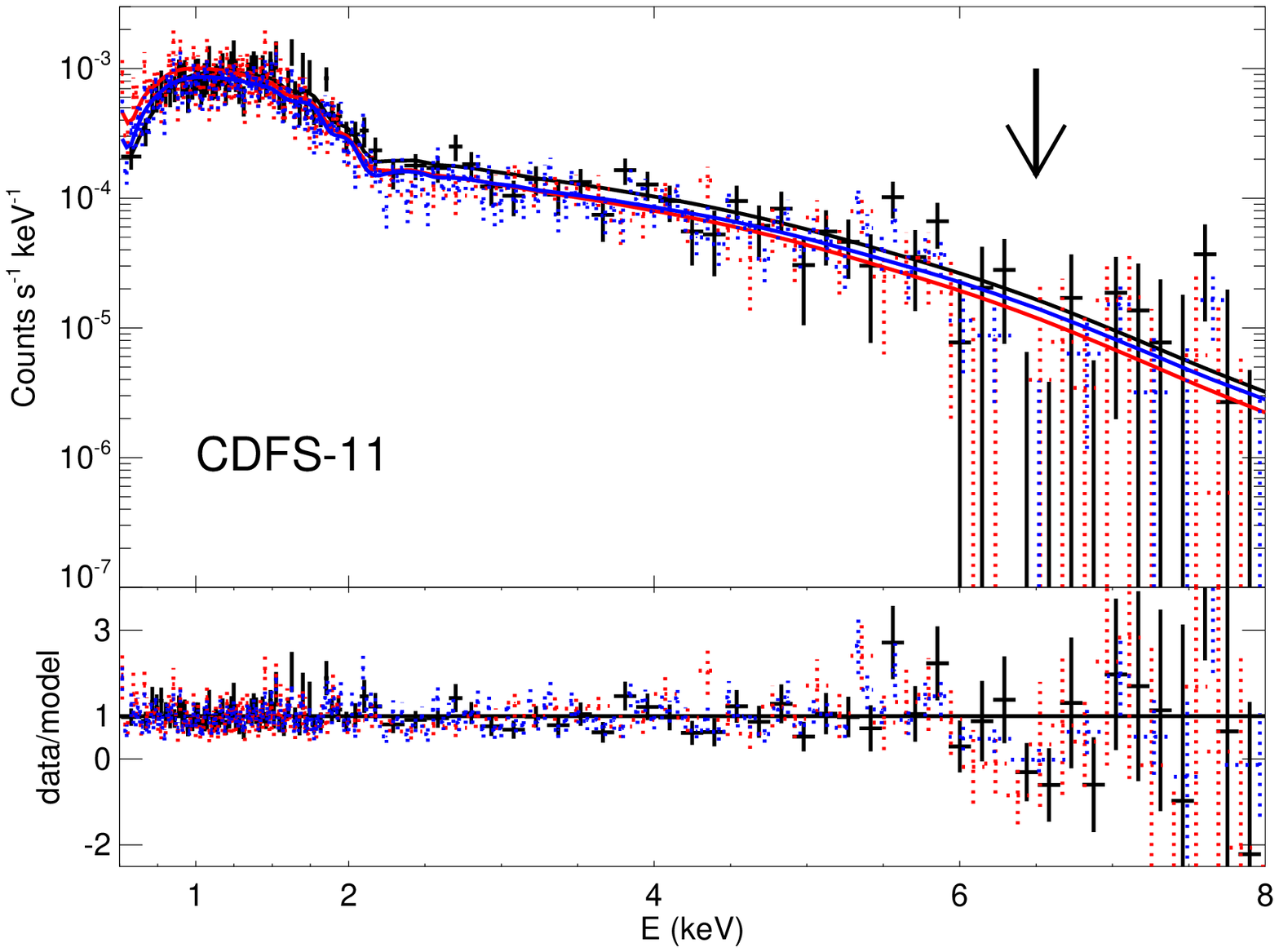}
\caption{
The new 1 Ms (black), old 1 Ms (red) and combined 2 Ms (blue) spectra of CDFS 11. The
unusual absorption line features at $\sim6.5$ keV (observed frame) are visible in all three spectra. The spectra were rebinned at least 10 channels per bin for display purpose only.
}
\label{spec11}
\end{figure}

The absorption feature at around 7 keV in CDFS 11 is also marginally
visible in the new 1 Ms exposure (see Fig. \ref{spec11}).
We added an absorption edge to our continuum model, but it did not improve
the fit significantly ($\Delta C \sim$ -2). We also tried to model the 
absorption feature by a saturated absorption line model. The model $notch$
of XSPEC was used by fixing the covering fraction at 0.99 to represent a 
blank absorption trough. The fitting was slightly better with $\Delta C = -4.9$ for 
two extra free parameters, yielding a confidence level of 93\% based on 
F-test. Note in the old 1 Ms spectra, the feature was detected with a
level of 99.98\%. This could probably be due to the significant higher
ACIS background in the new 1 Ms exposure. 

\section{Discussion}
\label{lineorigin}

Wang et al. (2003; 2005) have ruled out the possibilities that the unusual X-ray
features in CDFS 46 and CDFS 11 are due to any instrumental artifact, or
due to improper background subtraction. The conclusions hold for the new
1 Ms exposure, and get stronger since it's even more unlikely that any artifact
mentioned above, can affect both exposures obtained 7 years apart.
Statistical fitting results show that the confidence level of the unusual
features in both CDFS 46 and CDFS 11 is lower in the new 1 Ms exposure,
than early reported with old 1 Ms exposure. This is actually expected due
to the increasing \emph{Chandra} ACIS background with time.\footnote{The ACIS background rates remained relatively constant from the year 2000 until the end of 2003, and started to increase in 2004 with 10\% per year, independent of variability induced by the solar cycle. (See Section 6.16.2 of the Proposers' Observatory Guide v. 10 $-->$ http://cxc.harvard.edu/proposer/POG/) }
Note the blueshifted absorption feature in CDFS 11 is only marginally
visible in the new 1 Ms spectrum, and we are unable to tell whether it
varied based on current data. Except for providing marginally support to the
reality of the unusual absorption feature in CDFS 11 with new 1 Ms spectrum,
we are not discussing on
its nature more than Wang et al. (2005) did. In this paper we will
focus on CDFS 46.

Wang et al. (2003) has interpreted the X-ray emission line in CDFS 46 as 
relativistic blueshifted iron K line intrinsic to the quasar at $z$ = 1.617
(see below
for further discussion). We noticed that
Basu (2006) proposed a different scenario to interpret the optical and
X-ray spectrum of CDFS 46. He suggested that all the optical and X-ray
emission lines are blueshifted, resulting from an ejection. However, we
find this scenario is unacceptable due to several reasons. First, all the emission
lines were identified to have negative redshift, i.e., the observed optical
emission lines were identified as infrared lines in the rest frame. 
However, such extremely unusual galaxies with high blueshifts have never been 
reported. Furthermore, the identification of three optical emission
lines yields a redshift/blueshift scatter of 2.8\%, significantly larger 
than 0.8\% we obtained. The scatter (corresponding to 8400 km/s) is even
larger than the full width of the broadest emission line identified in CDFS46
(see Fig. 6 of Szokoly et al. 2004). Finally, the strong X-ray emission
line was classified as S K$\alpha$, which is supposed to be rather weak, and
has never been detected before.

It's interesting to note that in the GOODS-S ACS images, a nearby fainter
galaxy ($i$ = 24.7) is detected, 1.1\arcsec\ to the southwest of the quasar
CDFS 46. We note that the possibility to have a random galaxy brighter than
$i$ = 24.7 within 1.1\arcsec\ to certain known galaxy in GOODS-S is only 4\%,
suggesting that the two galaxies are likely at the same redshift.
VLT/FORS spectroscopy observations have classified the nearby fainter galaxy
as an emission
line galaxy at redshift 1.609 based on the detection of [OII] 3727, and
photometric measurements (Vanzella et al. 2008). This is within 0.5\%
to the redshift of CDFS 46 ($z$ = 1.617)\footnote{The redshift of CDFS 46
was identified
based on the detection of CIV, CIII] and MgII emission lines, see Wang et al.
2003; Szokoly et al. 2004.}. This
provides further evidence showing the redshift measurement of CDFS 46
is reliable. CDFS 11 is not covered by GOODS ACS images.

\subsection{The Nature of the Outflow}

Wang et al. (2003) has listed several possible explanations to the unusual
strong emission line in CDFS 46, including iron emission line from a 
relativistic outflow intrinsic to the quasar, strong iron absorption edge
due to a cold relativistic outflow, or intervening low-redshift ($z$ $\sim$ 0.034)
type 2 AGN. In the high spatial resolution GOODS ACS images, only
CDFS 46 at $z$ = 1.617 and a nearby galaxy at $z$ = 1.609 (1.1\arcsec\
apart) are visible within 3.5\arcsec\ radius. Also, although strong Fe 
K$\alpha$
line from neutral iron at 6.4 keV is not unusual in type 2 AGN, variation in
the line central energy is not expected, contrary to what find in CDFS 46.
These facts firmly rule out an intervening low-redshift type 2 AGN.

As Wang et al. (2003) stated, the emission line feature could also be 
statistically fitted by a blueshifted strong Fe absorption edge due to
cold outflow. 
Spectral fitting to the 2 Ms spectrum shows that a heavy absorption
(with $N_H$ = 3.2$^{+0.4}_{-0.3}$ $\times$ 10$^{24}$ cm$^{-2}$),
covering 99.92$\pm0.04$\% of the direct continuum, is required.
Statistically we can not rule out this model based on the 2Ms exposure,
however, we point out that, in addition to a Compton-thick cold outflow
with a velocity of 0.65c, such a model can only work under more 
extreme conditions.
These include that, the intrinsic X-ray luminosity of CDFS 46 needs to be
$\gtrsim$ 500 times larger (L$_{2-10keV}$ $\sim$ 3.0  $\times$
10$^{46}$ erg/s), which is way too luminous for its radio and optical
emission\footnote{The intrinsic X-ray to optical ratio $\alpha_{ox}$
obtained under this model is 0.2.};
the leaking X-ray emission (or scattered X-ray
emission) can not exceed 0.12\%; the cold outflow needs to slow down from
0.69c\footnote{The velocity of 0.74c (page L90, right column, 5th line) in Wang et al. (2003) should be corrected to 0.69c. There is a mistake in the calculation from Doppler factor to velocity in Wang et al. (2003).} to 0.65c after 2.7 years (in the quasar rest frame), and have to be locate
at a large distance from the unusual strong central continuum, to avoid to 
be ionized, otherwise we would see strong
soft X-ray emission which is free from absorption by an ionized absorber.
Assuming a density of 10$^{11}$ cm$^{-3}$, which is typical of the Broad Line Region in AGN,
we obtained a distance  $r$ $>$ $2\times10^{16}$ cm from the central emission.
Assuming the outflow cloud has a round or square shape, the outflow rate is then
$\dot{M}=4\pi f_c *r^2n*m_p*v_{wind}\simeq$ 
$(4\pi f_c)*2.1\times 10^4 M_{\sun}$ yr$^{-1}$, which is higher than
the Eddington accretion rate $\dot{M_E}=2.2$ M$_{\sun}$ yr$^{-1}$ of a
10$^8$ solar mass black hole, 
even for a very small covering factor of the outflow $f_c$ $\sim$ 10$^{-4}$. 

Therefor, the line feature we detected is most plausibly blueshifted
iron line due to a relativistic outflow intrinsic to the quasar.
The Doppler factor of 2.2 (for H- like Fe$_{XXVI}$ K$_\alpha$, 6.97 keV) 
implies bulk velocity of $\sim$0.65c (head-on) that must be 
responsible for the blueshift and relativistic boosting intensity.
Assuming a lower ionization state would yields higher velocity up-to
0.7c (for neutral Fe K$_{\alpha}$ line).
If the line is due to fluorescent emission or recombination emission
from plasma ionized by X-ray continuum, the continuum must illuminate
the outflow from sides to produce strong line with equivalent width
above several keV.
In this case, the equivalent width of the line can be boosted by a factor
of DF$^{3+\Gamma}$, where  DF means the Doppler factor for the bulk motion,
and $\Gamma$ the photon index of the X-ray continuum.

If the outflow is not ionized by the central X-ray emission,
considering the recombination timescale for Fe$_{XXV}$ is
$t_{recomb}\sim 3\times 10^4 Z^{-2}T_5^{1/2}n_9^{-1}$ s (Allen 1973, here T$_5$ is the temperature in unit of 10$^5$ K, and n$_9$ the electron density in unit of $10^9$cm$^{-3}$.),
typically in the range between 4.4$\times 10^3$ and 4.4 s
(Chartas, et al. 2002),
there must be some mechanics (such as magnetic driving) which keep heating the
relativistic outflow.
Migliari, Fender, \& Mendez (2002) had reported the discovery of blueshifted
and very strong iron emission lines (with velocity of 0.26c and equivalent width 
of 13 keV, Migliari, private communication)
 from extended X-ray emission in the X-ray binary system SS 433.
Their discovery implied the existence of large-scale reheating of baryon, 
which may work for the extreme blueshifted iron line in CDFS 46. 
We note that the new HETG observation of SS 433 also shows relativistic 
red- and blue- shifted lines from the central part indicating two side relativistic jets
(Marshall et al. 2002, Lopez et al. 2006). 
Taking the velocity of the
outflow as 0.70c, the distance which the outflow travelled during the
7 years' observation gap is $\sim0.6$ pc = 1.8$\times 10^{18}$ cm (in the quasar rest frame), which is $\sim$5 times the dust
sublimation radius ($R_d=0.4L_{45}^{1/2}\simeq$0.13 pc, here L$_{45}$ is the intrinsic 0.5-10 keV luminosity in unit of $10^{45}$ erg s$^{-1}$.). The distance is
too large for an adiabatic cooling process except for the existence of
reheating (Migliari, Fender, \& Mendez 2002).

The marginal change in the observed line central energy can either be due to a slow down
of the outflow (from 0.69c to 0.65c for H-like Fe K$\alpha$ line), or a change
of the ionization state, for example, from H-like Fe K$\alpha$
dominated to He-line Fe K$\alpha$ dominated (see table 3).
In the former case, we expect weaker emission line due to the weaker
boosting effect, which is consistent with observation (see Fig 1 and
table 3). In the later case, the line could either goes stronger or weaker,
depending on the exact ionization state.
We note that Chartas et al. (2007a) discovered 0.92-yr and 5.9-day
(rest frame) variability of the outflowing absorbing gas, with velocity of $\sim$0.1c (0.05c) and $\sim$0.4c (0.36c) due to He-like Fe K$_\alpha$ (H-like Fe K$_\alpha$) resonant absorption, 
in the large gravitational lensed BAL quasar PG 1115$+$80.

\subsection{Why So Rare?}
\label{rare}
Our two quasars are among the several AGNs which show X-ray outflows with 
velocity $\geq$ 0.4 c (see Table 1). 
Considering the large number of X-ray spectra obtained for quasars with advanced
X-ray telescopes such as XMM and Chandra, why there are only several sources
show relativistic outflowing features in their X-ray spectra?
We contribute this to the orientation and observational selection effect. 
The broad absorption line outflow in 
quasars, with velocity up-to 0.1c, is believed to have a sky coverage of 20\%.
We propose that relativistic outflow with velocity $\geq 0.4$c has
similar or smaller sky coverage, thus can only be detected in a small fraction of quasars.

The instrumental bandpass, such as \emph{Chandra} (0.3--10 keV) and 
\emph{XMM-Newton} (0.3--12 keV), also limits the detection of relativistic
outflowing features to certain redshift range. Assuming a Doppler factor 
$\geq$2 (or a bulk velocity of $\geq$0.6 c), only H- like Fe K$_\alpha$ 
line at redshift above 1.0 can be detected below 7 keV in the observed
frame (where \emph{Chandra} \& \emph{XMM} are sensitive to detect enough counts).
Furthermore, to confirm an emission or absorption line feature in an 
X-ray spectrum requires high number of detected photons (more than several
hundreds), thus they can only
be detected in luminous sources or by long enough exposure.

\subsection{Connection with jet or BAL Outflows in AGN}
We have stated above that the strong blueshifted emission line could be possibly
due to line emission from collimated jets. It's interesting to note that all
the three sources presenting blueshifted emission lines in Table 1 are radio-loud,
suggesting the blueshifted emission lines and jets are somehow connected.

\begin{figure}
\plotone{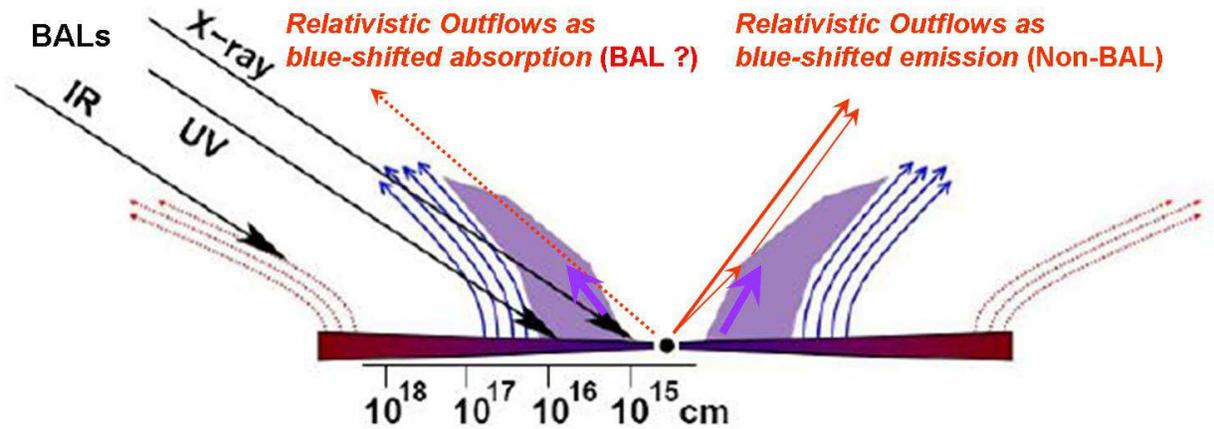}
\caption{ A diagram of the relativistic outflows (red lines and characters), based on the Figure 2 of Gallagher \& Everett (2007), which indicates the stratified outflows of radio-quiet quasars (the shielding gas is presented as solid shapes). We propose that an iron absorption line is produced when our line of sight is covered by the innermost shielding gas. In this case, the source does not necessarily appear as BAL in UV, as long as the X-ray shielding gas has larger sky coverage than the BAL region. An emission line is observed when our line of sight is not covered but close to the outflowing direction. }
\label{diag}
\end{figure}

Historically, in additional to jet, the most dominant examples of AGN
outflows have been found in Broad Absorption Line Quasars (BAL QSOs), 
which show absorption troughs in UV and optical lines with velocities
up to tens of thousands of km s$^{-1}$ 
(as large as 0.2 c, e.g., Hidalgo et al. 2007).
Disk wind model is one of most popular model to BAL outflows (see Proga 2007),
and the presence of X-ray shielding gas is required to prevent the gas from
over ionization and attenuate the X-ray continuum as observed (with
$N_H$ $\geq10^{23}$ cm$^{-2}$, Gallagher \& Everett, 2007).
We propose that the innermost region of the X-ray shielding gas could be
outflowing at a higher velocity (such as 0.6 $\sim$ 0.7 c), and more ionized,
which produces no photo-electric but He-like and H-like iron absorption to
central X-ray continuum.
Actually, Chartas et al. (2002; 2003; 2007a; 2007b) have identified blueshifted
X-ray absorption features in BAL QSOs at higher velocities and higher ionization
level ( see table 1).
An iron absorption line is produced when our line of sight is covered by the
innermost shielding gas, and an emission line is observed when our line of
sight is not covered but close to the outflowing direction 
(see Fig. \ref{diag}). Note the most recent simulation work of Sim et al. 
(2008) shows a similar and clear pattern that outflow can produce both iron 
absorption and emission features, depending on the viewing angle.
Taking into account the relativistic boosting effect, the blueshifted Fe K
emission line can be seen at a significant level. 
The innermost region of the X-ray shielding gas might have sky coverage
different from the broad absorption line region, thus the presence of 
X-ray absorption or
emission line features does not necessarily link to the presence of broad
absorption line in optical/UV band. However, it's worth to look for possible
blueshifted optical/UV features in CDFS 46 and CDFS 11, which requires optical
spectra with much higher quality. On the other hand, the search for blueshifted
X-ray features in high S/N X-ray spectra of BAL QSOs would also be helpful.

Finally, we note that BAL outflow could be aligned with jets, at least in some
BAL QSOs, and they might be physically connected (e.g. see Wang et al. 
2008). This suggest that both mechanisms (jet or BAL outflow) could probably work 
simultaneously to produce blueshifted emission/absorption lines with $v$ $>$ 0.4c
in the X-ray spectra of quasars.

\acknowledgments
The work was supported by Chinese NSF through NSFC10773010, NSFC10533050
and the CAS "Bai Ren" project at University of Science and Technology of China.
\clearpage


\begin{deluxetable}{lcccccc}
\tablecaption{A list of quasars showing relativistic Outflows with $v\geq0.4c$ in X-ray spectra.}
\tablecolumns{6}
\tablewidth{0pt}
\startdata
\hline
Name 	&	Type & $z$ & E$_{rest}$	&	Type & v$_{outflow}$ & Reference \\
	  	&		&	  &	(keV)	&		&  (c)	& \\\hline
$absorption$ & & & & & & \\
PG 1115$+$080 & mini-BAL RQQ& 1.72 & 7.3, 9.8 & Fe$_{XXV}$ K$_{\alpha}$ & 0.1, 0.4 & 1\\
H 1413$+$117 & LoBAL RQQ& 2.56  & 9, 15 & Fe$_{XXV}$ K$_{\alpha}$ & 0.23, 0.67 & 2  \\
APM 08279$+$5255 & BAL RQQ & 3.91 & 8.1, 9.8 & Fe$_{XXV}$ K$_{\alpha}$ & 0.2, 0.4 & 3 \\
CXO J033260-274748 & non-BAL RQQ & 2.579 & 22.2 & Fe$_{XXVI}$ K$_{\alpha}$ & 0.83 & 4 \\
$emission$ & & & & & & \\
CXO J033225-274219 & RLQ & 1.617 & 16.2 & Fe$_{XXVI}$ K$_{\alpha}$ & 0.69 & 5 \\
PKS 2149-306 & RLQ & 2.345 & 17.2 & Fe$_{XXVI}$ K$_{\alpha}$ & 0.71 & 6\\
PKS 0637-752 & RLQ & 0.654 & 1.60 & O$_{VII}$ K$_{\alpha}$ & 0.77 & 7
\enddata
\tablenotetext{NOTE.}{--Outflows with smaller ($v<0.4c$) relativistic velocities can be found at table 1 of Cappi 2006 and table 1 of Chartas 2007. Col.(1): galaxy name. Col.(2): galaxy type - BAL: broad absorption line; LoBAL: low-ionization BAL; RQQ: radio-quiet quasar; RLQ: radio-loud quasar  Col.(3): redshift Col.(4): Line rest-frame energy in units of keV Col.(5):Fe transition type Col.(6): Velocity (in units of c). Col (7): References - 1) Chartas et al. 2003, 2007a; 2) Chartas et al. 2007b; 3) Chartas et al. 2002, 2007a, Hasinger et al. 2002; 4) Wang et al. 2005 and this paper; 5) Wang et al. 2003 and this paper; 6) Yaqoob et al. 1999; 7) Yaqoob et al. 1998.}
\end{deluxetable}
\clearpage

\begin{deluxetable}{l|c|cc|c}
\tablecaption{Spectral fitting to CDFS 46}
\tablecolumns{6}
\tablewidth{0pt}
\startdata
\hline
		& 	New 1 Ms	& Old 1 Ms & Wang03	& Comb. 2 Ms  \\   \hline
N$_H$(10$^{22}$cm$^{-2}$)  & $<$ 0.3 & 1.2$^{+0.6}_{-0.8}$ & 1.4$^{+0.4}_{-0.3}$  & 1.1$^{+0.6}_{-0.6}$\\
$\Gamma$ & 1.8$^{+0.2}_{-0.1}$ & 2.4$^{+0.3}_{-0.3}$ & 2.2$^{+0.2}_{-0.1}$ &2.0$^{+0.2}_{-0.2}$ \\  
C/dof(continuum only)	&  252.1/287 &  241.0/246 & 571.4/578 & 348.6/372 \\ 
E$_c$ (keV)& 5.82$^{+0.16}_{-0.13}$ & 6.2$^{+0.1}_{-0.1}$ & 6.2$^{+0.2}_{-0.1}$  &  6.1$^{+0.3}_{-0.2}$ \\ 
$\sigma$ (keV) & 0.15$^{+0.25}_{-0.15}$ & 0.18$^{+0.17}_{-0.08}$ & 0.2$^{+0.2}_{-0.1}$ & 0.3$^{+0.3}_{-0.2}$ \\   
I 	& 0.9$^{+0.6}_{-0.4}$ & 1.4$^{+0.7}_{-0.6}$ & 1.3$^{+1.0}_{-0.4}$ & 1.3$^{+0.8}_{-0.6}$\\ 
EW (keV) & 1.9$^{+1.3}_{-0.8}$  & 5.7$^{+2.9}_{-2.4}$ & 4.4$^{+3.2}_{-1.4}$  & 3.6$^{+2.2}_{-1.7}$ \\
C/dof(continuum+line)	& 233.9/284	& 216.4/243 & 548.8/575 &  317.7/369 \\
F test	& 99.99\%	& 99.999\% & 99.996\%	& $>$99.9999\%   \\\hline
\enddata
\tablenotetext{NOTE.}{--Here dof stands for degree of freedom. The two lines of "C/dof" are statistical results without and with emission line. The line intensity $I$ is in units of 10$^{-7}$photons cm$^{-2}$s$^{-1}$. The emission line
parameters are given in the observed frame. 
The new 1 Ms, old 1 Ms and combined 2 Ms spectra are binned to have at least 
one count per bin. We also quote the fitting results from Wang et al. (2003)
for the old 1 Ms unbinned spectrum, which are consistent with those from the
binned spectrum.
} 
\end{deluxetable}

\clearpage


\begin{deluxetable}{cccc}
\tablecaption{Two possibilities for the variability of the emission feature in CDFS46}
\tablecolumns{6}
\tablewidth{0pt}
\startdata
\hline
a: H- like Fe K$_{\alpha}$, && assuming $\theta = 0 $. & \\\hline
Year  & F$_{intri}^a$ & EW$_{intri}^b$  & v$_{bulk}$/c \\ 
2001      & 4.4$^{+3.4}_{-1.4}$ &59$^{+30}_{-25}$  & 0.69$^{+0.01}_{-0.01}$  \\
2007      & 4.0$^{+2.6}_{-2.2}$ &44$^{+29}_{-24}$  & 0.65$^{+0.02}_{-0.01}$  \\
\hline
b: H- like Fe K$_{\alpha}$&to He- like Fe K$_{\alpha}$,& assuming $\theta = 0 $. &  \\\hline
Year & F$_{intri}^a$ & EW$_{intri}^b$  & v$_{bulk}$/c \\ 
2001      & 4.4$^{+3.4}_{-1.4}$ & 59$^{+30}_{-25}$  & 0.69$^{+0.01}_{-0.01}$  \\
2007      & 3.4$^{+2.4}_{-1.9}$ & 27$^{+18}_{-12}$  & 0.67$^{+0.02}_{-0.01}$ 
\enddata
\tablenotetext{a}{The unit of the F$_{intri}$ is $10^{-9}$photons cm$^{-2}$ s$^{-1}$.}
 \tablenotetext{b}{The unit of the EW$_{intri}$ is eV.}
\end{deluxetable}

\end{document}